\documentclass[letter]{aa} 
\usepackage{graphicx}
\usepackage{natbib}
\usepackage{txfonts}
\begin{document}
   \title{Chromospheric activity as age indicator}

   \subtitle{An L-shaped chromospheric-activity versus age diagram.}

   \author{G. Pace
          \inst{1}
          }

   \institute{Centro de Astrof\'{\i}sica, Universidade do Porto,
              Rua das Estrelas, 4150-762 Porto, Portugal\\
              \email{gpace@astro.up.pt}
             }

   \date{Received September 15, 1996; accepted March 16, 1997}

 
  \abstract  {Chromospheric activity  has been  calibrated  and widely
    used as an age indicator.  However, it has been suggested that the
    viability of this  age indicator is, in the  best case, limited to
    stars younger than about 1.5 Gyr.}  {I aim to define the age range
    for which chromospheric activity is a robust astrophysical clock.}
             {I  collected literature measurements  of the  S-index in
               field stars, which is a  measure of the strength of the
               H  and K  lines of  the  Ca {\rm  II} and  a proxy  for
               chromospheric  activity, and exploited  the homogeneous
               database  of  temperature  and age  determinations  for
               field stars provided by the Geneva-Copenhagen survey of
               the solar neighbourhood.}  { Field data, inclusive data
               previously   used  to  calibrate   chromospheric  ages,
               confirm  the  result  found  using open  cluster  data,
               i.e. there is no  decay of chromospheric activity after
               about 2 Gyr.}  {The only existing indication supporting
               the viability of chromospheric ages older than 2 Gyr is
               the similarity of  chromospheric activity levels in the
               components of  35 dwarf binaries. However,  even in the
               most   optimistic    scenario,   uncertainty   in   age
               determination  for field stars  and lack  of sufficient
               data in  open clusters make any  attempt of calibrating
               an age  activity relationship for  old stars premature.
               The hypothesis that  chromospheric activity follows the
               Skumanich  law,  i.e.    that  it  is  proportional  to
               t$^{-1/2}$, should be relaxed.
\footnote{Table 1 is only available  in electronic form at the CDS via
  anonymous   ftp  to   cdsarc.u-strasbg.fr   (130.79.128.5)  or   via
  http://cdsweb.u-strasbg.fr/cgi-bin/qcat?J/A+A/} }


   \maketitle
%

\section{Introduction}

Chromospheric activity  (CA) is  the amount of  energy emitted  in the
chromosphere,  i.e.    the  layer  of  a  late-type   star  above  the
temperature gradient inversion. This energy is radiated especially
through emissions of strong resonance lines, whose flux is therefore a
proxy  for CA.   The  S index  is a  measure  of the  strength of  the
chromospheric emission core  of the H and K lines of  the Ca {\rm II},
F$^{\prime}_{HK}$ is  its conversion into  flux, and R$^{\prime}_{HK}$
is the flux  normalized to the bolometric emission  of the star. These
are  the  most  commonly used  proxies  for  CA,  and they  have  been
monitored for decades  in nearby stars in the  Mount Wilson \citep[see
  e.g.][]{baliunas95}   and,   more   recently,   in   several   other
observational campaigns \citep[see e.g.][]{hall07}.

The CA  declines during the stellar  life time, as  does the rotation,
which powers  it. Therefore  CA is in  principle a  good astrophysical
clock and, as such, has  been calibrated and widely used. However, CA,
on top  of its long-term  decay, undergoes short-term  variations. The
Sun,  for instance, has  an activity  cycle of  11 years with  secular
variations that caused  the Maunder minimum.  This sets  limits on the
precision of chromospheric ages based on one-epoch CA measurements.

Observations  in open  clusters suggested  that  CA in  stars in  open
clusters at about 1.5 Gyr, such as IC4651 and NGC3680, already dropped
to  solar levels \citep{pace1,  lyra}.  This  result was  confirmed by
\cite{zhao} who studied wide binaries.  \cite{pace3} claimed that even
in the age range between 0.7 and 1.2 Gyr any CA evolution must be less
important than  short-term variations,  and that a  sudden drop  of CA
occurred  on a  very  short time-scale:  0.2  Gyr.  These  conclusions
relied on the age determinations of \cite{salaris}. In fact, according
to these, the  ages of the active clusters,  which all spanned roughly
the  same interval  of  CA levels,  ranged  from 0.7  Gyr (Hyades  and
Praesepe) to 1.2 Gyr (NGC5822),  while the ages of NGC3680 and IC4651,
which are  as inactive as  the Sun, were  estimated to be 1.4  and 1.7
Gyr,   respectively.    A  more   recent   set   of  homogeneous   age
determinations   for   these   clusters   makes   NGC5822   0.9   Gyr
\citep{carraro11},  IC4651   1.5  Gyr,  and  NGC3680   1.75  Gyr  old
\citep{anthonytwarog09}.  Therefore, according  to the CA measurements
of \citeauthor{pace3} and these new age determinations, the transition
from chromospherically active to inactive would occur in 0.6 Gyr, and,
most important, the interval of time during which CA remains virtually
constant is reduced to 0.2 Gyr.  In other words, it is still warranted
to investigate the viability of CA as an age indicator until 1.5 Gyr.

In  contrast   to  the   result  of  \cite{pace1},   \cite{lyra},  and
\cite{zhao}, several calibrations of CA  as an age indicator have been
published   \citep[see  e.g.][]{soderblom91,donahue93,lachaume99,mh08}
based  on field stars,  binary systems,  or open-cluster  members, and
these calibrations are claimed to be valid at all ages.

The problem that all of these studies face is the paucity of good data
for open  clusters older  than about  1 Gyr, M67  being the  only such
cluster with  a rich sample  of high-quality CA measurements  made for
stars spanning  a wide range of  temperatures \citep{giampapa06}.  The
sample of  \cite{pace1} and \cite{pace3}  include, in addition  to M67
members, only seven stars older than  1 Gyr, two dwarfs in NGC3680 and
five in IC4651, and they are all hotter than the Sun.

\cite{mh08} reported only two more open clusters older than 1 Gyr with
CA  measurements:  NGC752 (15  stars  observed)  and  NGC188 (3  stars
observed)  originally observed  by \cite{barry87}  and \cite{barry84}.
However, these data were considered suspicious \citep{soderblom91} and
\cite{mh08}  did not  use them  either in  their Table  5 or  in their
Fig. 4, where other cluster data are reported.

Clearly,  more  open-cluster  data  are needed.   However,  since  the
publication of  the most  recent CA calibration  I have  knowledge of,
more   S-index  measurements   were  published,   and   new,  precise,
homogeneous determinations  of stellar  ages and temperatures  for the
solar  neighbourhood have  been  made available.   This  allows us  to
reexamine CA evolution through the study of field stars.

\section{The data}
\label{sec:data}

\subsection{Field data}
\label{sec:field}

All   the  available   data  in   the  literature   reporting  S-index
measurements for field  dwarf stars in the same  scale as Mount-Wilson
data  were  collected.   The   data  include  the  following  sources:
\cite{arrigada11MagellanPSP},                    \cite{jenkins11FEROS},
\cite{baliunas95},                                \cite{knutson10TPHS},
\cite{lopez10active_nearby_stars},                 \cite{gray03NSTARS},
\cite{schroder09_rapidrot},                     \cite{cincunegui07Arg},
\cite{lockwood07MWandlowell},   \cite{duncan91},  \cite{gray06NSTARS},
\cite{tinney02AngloAustralianPS},                        \cite{hall07},
\cite{white07FEPSstars},    \cite{henry96},   and   \cite{wright2004}.
\cite{strass} measured  absolute H and K emission-line  fluxes of more
than  1000   late-type  stars,  but   they  did  not   report  S-index
measurements.   I  transformed   their  flux  measurements  into  line
strength (HK$_{strass}$)  by reverting their equations 2)  and 3), and
then to  an equivalent of  the S-index in  the Mount Wilson  scale, by
calibrating a  relationship between HK$_{strass}$ and  the S-index for
the 329 stars in their sample in common with at least one of the other
works mentioned above. The spread around a linear relationship between
HK$_{strass}$ and S-index is roughly  10 \%, compatible with CA cycles
and  errors. I  also included  data from  \cite{buccino08},  which are
measurements  of the  strength  of the  Mg  {\rm II}  h  and k  lines,
transformed into the S index through  a calibration made with a set of
117 nearly  simultaneous observations of Mg  {\rm II} and  Ca {\rm II}
fluxes  of 21  main-sequence stars.   Data from  Lockwood et  al.  and
Baliunas et al.  are taken from  the Mount Wilson survey and cover the
entire  activity  cycle  for  each  target. I  adopted  them  whenever
possible.  For  the other  stars, when more  than one  measurement was
available  in the  literature, the  adopted S-index  was taken  as the
average  between the  highest  and the  lowest individual  measurement
recorded, which  best represents  the time-averaged CA  level, because
typically  stars   are  not  observed  during  an   entire  CA  cycle.
Unfortunately,  some authors  did not  report single  measurements but
only  indicated the average  S-index.  \cite{hall07}  reported S-index
measurements only for part of their  total sample, for the rest of the
stars  only R$^{\prime}_{HK}$  values are  given, and  the photometric
information they  used to transform S-index  into R$^{\prime}_{HK}$ is
not indicated either.   This problem affects 27 stars  of their sample
that could have been used  otherwise.  Only one of these, HD70110, was
not measured in any of the other aforementioned works.

Although  S-index  measurements  come  from  a  variety  of  different
sources, they  were all calibrated to  match the same  scale, which is
the Mount  Wilson scale. The most  significant source of  error is the
intrinsic  variability  of CA.   The  conclusions  presented here  are
independent  of the  choice  of the  sources,  i.e.  they  hold if  we
discard one or more of the studies from the compilation, provided that
sufficient data are analysed.

To  guarantee homogeneity  of  temperature and  age evaluations,  this
information was  taken from the Geneva-Copenhagen survey  of the Solar
neighbourhood (GCS): a set of determinations of metallicity, rotation,
age,    kinematics,   and    Galactic   orbits    for    a   complete,
magnitude-limited, and kinematically unbiased sample of 16682 nearby F
and G dwarf  stars \citep{GCSI}. Two revised versions  of the GCS were
published  by  the  same  group  based on  improved  uvby  calibration
\citep{GCSII}   and   on   revision   of  the   Hipparcos   parallaxes
\citep{GCSIII}.   I used  the last  revision by  \cite{GCSluca}, which
adopts  the infrared  flux method  (IRFM) to  improve the  accuracy of
temperature estimations.   For this  method, with the  most up-to-date
calibration, the uncertainty in the determinations of the $T_{eff}$ is
of the  order of 30 K,  plus a systematic error,  i.e.  uncertainty in
the zero point, of 15 K \citep{casagrande06, casagrandecal}.  Accurate
information on binarity is also given in this last version of the GCS,
and I used it to select single stars.

By cross-correlating  the collection of S-index  measurements with the
GCS data set and discarding  multiple systems, I finally selected 1744
single dwarves.  The data collected about these stars are available at
the  CDS in  electronic  form.  From  these,  the 494  stars with  age
estimations  precise enough  for the  scope of  the present  work were
additionally       selected       ($\sigma_{age}<$2       Gyr       or
$\frac{\sigma_{age}}{age}<$0.3).

S-indices  were  transformed   into  R$^{\prime}_{HK}$  following  the
procedure of \cite{noyes84}, which  involves B-V colours.  The adopted
values for B-V  where obtained from the temperatures  by reverting the
calibration  of \cite{casagrandecal}.

\subsection{Open cluster data}
\label{sec:dataoc}

I used here log~R$^{\prime}_{HK}$ evaluations and B-V colours from the
compilation   of  \cite{mh08}   for  the   Ursa  Major   moving  group
(Collinder~285),  Hyades, and  M67, while  I ignored  data  on younger
clusters of the same source, since they are not relevant for the scope
of  the  present  work.   I  added  data of  NGC752  and  NGC188  from
\cite{barry87}, and  from \cite{pace3}  the data of  Praesepe, IC4756,
NGC5822, IC4651,  NGC3680, as well as  one Hyades member  and four M67
members   not   included   in   the   sample   of   \citeauthor{mh08}.
\citeauthor{pace3}  used spectroscopic  temperatures for  their stars,
and I adopted them here.  The other temperatures were obtained through
the  calibration of \cite{casagrandecal}  from B-V  dereddened colours
given in the respective source.

The S-index measurements of  \citeauthor{barry87} were made on spectra
at  a   lower  resolution  ($\sim$2{\AA}  FWHM)   than  the  classical
Mount-Wilson data,  but they were converted to  match the Mount-Wilson
system.  However, when I transformed them into R$^{\prime}_{HK}$ using
\cite{noyes84},  I obtained results  completely inconsistent  with the
values published in  \citeauthor{barry87}, with typical differences of
0.1 in logarithmic scale, and even larger when taking into account the
correction suggested  by \cite{soderblom91}.  I believe  that Barry et
al.'s  S-index measurements  are trustworthy,  since  their conversion
into R$^{\prime}_{HK}$  gives reasonable results, while  both Barry et
al.'s  values  of  R$^{\prime}_{HK}\times10^5$  and  their  correction
following  \cite{soderblom91}  are  suspicious.   As  for  the  former
estimations, in addition to being inconsistent with the S-indices from
which    they   are    supposed   to    be   derived,    the   highest
log~R$^{\prime}_{HK}$  value  for NGC752  exceeds  that  for the  most
active  star among  Hyades and  Collinder~285  by 0.38  dex.  This  is
highly  implausible since  Hyades and  Collinder~285 are  younger than
NGC752 and their sample considered here outnumbers by more than 3 to 1
that    of    NGC752.     By    introducing    the    correction    of
\citeauthor{soderblom91}, this probably spurious value is decreased by
only   0.01  dex   in   the  logarithmic   scale,   and  one   obtains
log~R$^{\prime}_{HK}$=-5.41  dex for  the  least active  of the  three
stars of NGC188.  This is not  quite as suspicious as the case for the
most active  star in  NGC752, but still  calls for some  mistrust.  In
fact, it is 0.28 dex lower than the log~R$^{\prime}_{HK}$ value of the
least  active of  the 26  stars in  M67. In  addition, among  the 1744
single stars with  CA measurements collected here, only  12 stars have
log~R$^{\prime}_{HK}<$-5.41 dex, while  the most active NGC188 member,
with \citeauthor{soderblom91}'s correction, would  have a CA above the
average  of field  stars.  I  therefore used  R$^{\prime}_{HK}$ values
derived  from published S-indices  for NGC752  and NGC188.   Because I
failed   to    identify   the   star   numbering    system   used   by
\citeauthor{barry87}, and they did not report the stellar coordinates,
I could only  use the B-V colours given in  the paper without updating
the  photometry and  membership  information. The  CA measurements  in
NGC752 and NGC188  should be taken with a grain  of salt.  The S-index
measurements from  \cite{pace3} are based  on high-signal-to-noise and
high-resolution  spectra taken  with UVES  at  the VLT,  but they  are
performed in  a different way, using  only the K-line.  I  used the 17
stars, either  M67 or  Hyades members and  the Sun, in  common between
\cite{pace3} and  the compilation of  open clusters by  \cite{mh08} to
find   a  linear   relationship  between   the  S-index   measured  by
\citeauthor{pace3}  and the  Mount-Wilson  one, and  to transform  the
former into the Mount-Wilson scale.  The 17 stars in common span large
part of  the entire  CA range  of the open  cluster sample,  and their
deviation  from the  the linear  regression is  of only  $\sim  5 \%$,
therefore the transformation is dependable.

\section{Results and discussion} 
\label{sec:results}

\subsection{Field stars}
\label{sec:results:field}

   \begin{figure}
   \centering
   \includegraphics[width=8cm]{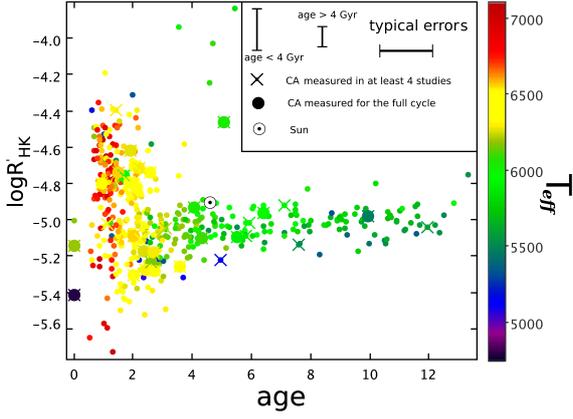}
      \caption{log~R$^{\prime}_{HK}$ versus  age diagram.}
         \label{fig:1}
   \end{figure}
%

In  Fig.   \ref{fig:1}  the  sample  of  494 stars  is  plotted  in  a
log~R$^{\prime}_{HK}$  versus  age  diagram.   Typical  errors  in  CA
reported  in  the figure  are  estimated by  looking  at  the full  CA
variations for  the stars measured twice  or more.  For  about 95\% of
those stars,  full variations are within  0.4 dex, and  within 0.2 dex
for stars  older than 4  Gyr, therefore I  adopted 0.2 dex  as typical
error for  stars younger  than 4 Gyr  and 0.1  dex for the  others.  1
$\sigma$ errors in  age were imposed, as a  selection criterion, to be
either less than 2 Gyr or less than 30\% of the age.  As a result, the
typical  horizontal error  bar  is  of about  1.5  Gyr.  The  L-shaped
distribution  of the  bulk of  the  data points  in Fig.   \ref{fig:1}
strongly suggests that the decay  of CA stops completely at a relative
early age. This age may appear  to be about 3 Gyr in Fig.  \ref{fig:1}
as a result  of the errors that scatter  the data points horizontally,
but  could  be  much  younger  in  reality.  It  is  not  possible  to
understand how CA evolves in this first phase, because the uncertainty
in the  age is of the  same order of  magnitude as the time  span over
which the  decay occurs.  No matter  how data are  further selected on
the  basis of  the  number of  CA  measurements available  and on  the
temperature, there  is no indication  of a CA-age  anticorrelation for
older stars.  The vertical spread  of the points relative to old stars
is  compatible with being  caused by  cyclic variations,  which proves
that  CA  measurements are  reliable  also  for  inactive stars.   The
presence of very  inactive stars only at young  ages is connected with
the  temperature.   Indeed,  among  the  1744  single  stars  with  CA
measurements collected here,  the 11 least active are  all hotter than
6300 K, and are therefore young.

Following   \cite{meteffect},  I   checked  the   correlation  between
metallicity   and  log~R$^{\prime}_{HK}$:   the   Pearson  correlation
coefficient between them is roughly -0.2.

\subsection{Open clusters}
\label{sect:resultoc}

\begin{figure}
\centering
\includegraphics[width=9cm]{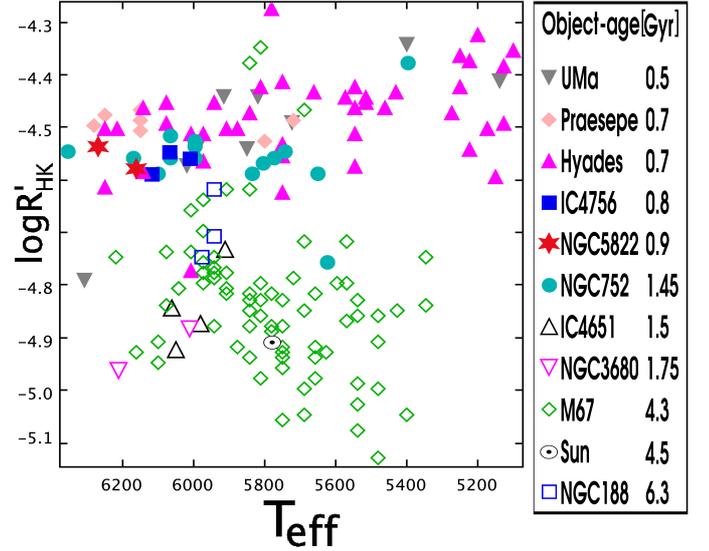}
\caption{ log~R$^{\prime}_{HK}$ versus  dereddened B-V colour for open
  clusters and the Ursa Major  moving group (Collinder~285, UMa in the
  legend).}
\label{fig:2}
\end{figure}

Fig.   \ref{fig:2} shows  a  log~R$^{\prime}_{HK}$ versus  temperature
diagram.  The ages reported in the legend are taken from the following
sources   (in    order   of   adopted    priority):   \cite{carraro11,
  anthonytwarog09, salaris, king03}.

Data  points in  Fig.  \ref{fig:2}  can  be broadly  divided into  two
groups,  one with  the bulk  of the  young-cluster members  (age $\leq
1.45$  Gyr,  filled symbols),  with  a  slightly  increasing trend  of
log~R$^{\prime}_{HK}$  with decreasing temperature,  and one  with the
bulk  of old  cluster members  (age  $\geq 1.5$  Gyr, empty  symbols),
characterized by  a more complicated  pattern of log~R$^{\prime}_{HK}$
as a  function of  temperature. Only  three of the  78 members  of the
former group are less active than log~R$^{\prime}_{HK}$=-4.63 dex, and
only three out of the 89 members of the latter group, all M67 members,
are more active than  log~R$^{\prime}_{HK}$=-4.60 dex.  The reason why
three M67 stars  are at a Hyades activity  level is probably binarity,
as confirmed by the fact that one of these stars is Sanders~747, whose
spectrum taken with UVES at  the VLT (observing run 66.D-0457) clearly
shows that it is a double-lined  binary.  It follows that no more than
$\sim$ 5\% of the single stars  should be expected to have an activity
atypical for  their age.   Although the CA  levels spanned by  the two
groups almost overlap, the average difference between them is as large
as $\Delta$log~R$^{\prime}_{HK} \sim$ 0.4 dex, i.e.  twice the typical
error for young stars and four times as much that for old stars.

According to the  most recent age determinations, the  only cluster of
the active group significantly older than the Hyades studied so far is
NGC752,  which is  on average  slightly  less active  than Hyades  and
Collinder~285, even  though all NGC752 members, with  the exception of
one outlier, have  CA levels within the range  encompassed by the bulk
of   Hyades    stars.    The   drop   of   the    average   value   of
log~R$^{\prime}_{HK}$  between  the age  of  NGC752  and  that of  the
youngest inactive  cluster is  $\sim$ 0.3 dex  or more.   The youngest
inactive cluster, according to \cite{anthonytwarog09}, is IC4651, only
50  Myr older  than NGC752.   Therefore,  even after  the revision  of
open-cluster       ages       by      \citeauthor{carraro11}       and
\citeauthor{anthonytwarog09}, we  still have  an indication of  a very
rapid CA  drop.  However, the dependability  of the data  on NGC752 is
questionable (see Sect. \ref{sec:dataoc}).

The dependability of the NGC188 data is also questionable, and we have
only three  stars in  a narrow temperature  range. It  is interesting,
however,  that it  is  the oldest  cluster,  by far  more active  than
IC4651, NGC3680, and  M67, thus definitely pointing to  the absence of
CA decay after 1.5 Gyr.

\subsection{Accuracy of the measurements. }

Data  shown  in  Fig.   \ref{fig:1}  and Fig.   \ref{fig:2}  are  both
characterized by  the lack of old  and active stars,  as expected. The
presence  in Fig.   \ref{fig:1} of  young  and inactive  stars can  be
explained by the  errors in age.  In Fig.   \ref{fig:1}, the age limit
after which no  more active stars are present is 3  Gyr, while in Fig.
\ref{fig:2} this age limit is 1.5 Gyr.  Again, this is compatible with
the 1 $\sigma$ errors in  the age quoted by \cite{GCSluca}, which, for
the  upper-left  bunch   of  data  points  of  the   diagram  in  Fig.
\ref{fig:1}, peak  around 0.2 Gyr but  extend all over  to the imposed
upper limit  of 2  Gyr. There  is a significant  difference in  the CA
limit that sets the distinction between active and inactive stars: for
open  clusters this limit  is  at log~R$^{\prime}_{HK}$=-4.6,  while
looking  at the  field data,  the same  limit should  be the  solar CA
level. This  last circumstance seems  to be incompatible with  the Sun
being in the middle of the CA distribution of M67 members. This raises
the question  whether the OC  data are on  a scale different than the
field data, but this would  be surprising, since both \cite{pace3} and
\cite{giampapa06} found that  most of M67 members have  CA levels that
fall   in  the   range  spanned   by  the   Sun,  they   both  used
high-resolution and good signal-to-noise spectra  of the Ca II H (only
Giampapa et  al.) and K  lines, and they have  temperature estimations
that   perfectly  match  for three  solar-type  stars, but  differ
systematically by about 100 K for the other (hotter) stars.

\section{Summary and conclusions}
\label{sec:conclusions}

I collected a very large sample of published S-index measurements that
were calibrated to match the Mount-Wilson scale.  For field stars, I
used information on  temperatures and ages of the  very exhaustive and
homogeneous determinations  of the GCS, adopting the  last revision by
\cite{GCSluca}.

Not only  open-cluster data, but also field-star  data strongly favour
the conclusion of \cite{pace1}, \cite{lyra}, and \cite{zhao}: there is
no evolution of chromospheric activity after $\sim$ 1.5 Gyr.

\cite{mh08} collected  data on  35 dwarf binaries.   The CA  levels of
their components are strongly correlated, even if we consider only the
systems  more inactive than  log~R$^{\prime}_{HK}$=-4.60. This  is the
only piece  of evidence  supporting a correlation  between CA  and age
holding for old stars.  The  search for an alternative explanation for
this fact, invoking processes typical of binary systems, is definitely
warranted,  but the  possibility  that  the CA  is  a tight,  strictly
monotonic  function of  time, even  though extremely  unlikely  in the
light of the analysis presented  here, cannot be completely ruled out.
According to the present analysis, this function, if it exists, is not
like $\propto t^{-1/2}$,  as tentatively suggested by \cite{skumanich}
on  the basis  of very  scanty  data and  widely used  ever since  its
pubblication, and our present knowledge of it is too poor for it to be
an astrophysical clock of any use.

At present, the only prediction on age we can infer from CA is whether
a star  is younger  or older  than the open  cluster NGC752,  which is
estimated to be  1.45 Gyr old, with the caveat  that there exist young
and very inactive stars hotter than 6300 K. The situation is likely to
improve  rapidly  due  to  the  accumulation  of  asteroseismic  data.
Indeed, asteroseismology has proved  to allow age determination with a
good precision \citep{age1, age2, age3}.

\begin{acknowledgements}
\end{acknowledgements}

It is  a pleasure  to acknowledge Simone  Recchi and  Jos\'{e}-Dias do
Nascimento  Jr  for  useful  discussions.  The  referee's  suggestions
considerably improved  the quality of this manuscript.   The author is
supported  by  grant  SFRH/BPD/39254/2007 and  acknoweldges  financial
support  by  the  project  PTDC/CTE-AST/098528/2008,  both  from  FCT,
Portugal.

\end{document}